\documentstyle[12pt,epsfig]{article}

\makeatletter                    
\@addtoreset{equation}{section}  
\makeatother                     

\newskip\humongous \humongous=0pt plus 1000pt minus 1000pt

\newif\ifdtup

\def\a{\alpha}

\begin{document}
\title{Financial Modeling and Option Theory with the \\ Truncated Levy
Process}

\author{Andrew Matacz\thanks{ Email: andrewm@maths.usyd.edu.au}\\
{\small School of Mathematics and Statistics}\\
{\small University of Sydney, 2006} \\
{\small Australia}}
\date{\small {\it Report 97-28, October 1997}}
\maketitle
\begin{abstract}

In recent studies the truncated Levy process (TLP) 
has been shown to be very promising for the modeling of financial 
dynamics. In contrast to the Levy process, the TLP 
has finite moments and can account for both the previously observed
excess kurtosis at short timescales, along with the slow convergence to 
Gaussian at longer timescales.
I further test the truncated Levy paradigm using high frequency data from 
the Australian All Ordinaries share market index. 
I then consider, for the early Levy dominated regime, the issue of option 
hedging for two different hedging strategies that are in some sense optimal. 
These are compared with the usual delta hedging approach and found to differ
significantly. 
I also derive the natural generalization of the Black-Scholes option pricing 
formula when the underlying security is modeled by a geometric TLP. This 
generalization would not be possible without the truncation.

\end{abstract}
\newpage
\section{Introduction}
It has been widely appreciated for some time that fluctuations in financial 
data show consistent excess kurtosis indicating the presence of large 
fluctuations not predicted by Gaussian models. With the continuing growth in 
the derivatives industry, and the recent emphasis on better risk management 
practices, the need for models that can describe these large events has 
never been greater. 

Several authors \cite{levy-empirical} have explored the stable Levy class of 
distributions as a possible alternative to the Gaussian. Levy distributions 
exhibit scaling or fractal properties which occur commonly in complex systems
often studied in statistical physics.   
Indeed the Levy distribution 
does seem to provide a consistently better representation of financial data than 
the Gaussian distribution. Despite this it would be fair to say that there 
has been a lack of interest in this class of models, especially in the  
option theory literature (see \cite{levy-options} for some exceptions). 
One reason for this 
would have to be a reluctance to accept the infinite variance which is typical 
in these models. This makes it difficult to find an appropriate generalization 
of the firmly entrenched Black-Scholes option pricing framework.
Perhaps a more important reason is that financial data tends 
to become more Gaussian over longer timescales \cite{ak,ms}. This property 
is also evident in the decay of the implied volatility smile obtained from 
observed option prices with increasing maturity \cite{hob}. 
These properties cannot be 
explained by Levy distributions due to their stable additive property 
(the central limit theorem does not apply due to their infinite variance). 

A much more popular approach to explaining the excess kurtosis is based on 
the observation that the variance (or volatility in financial language) 
of financial data appears to behave randomly. Stochastic volatility will 
generate kurtosis in an otherwise Gaussian process and this has led to a body 
of literature that attempts to model volatility as a diffusion process 
(for a review see \cite{hob}). An important reason for the popularity of 
this approach is that it is still based on a Gaussian framework making 
it a relatively Black-Scholes friendly explanation for excess kurtosis. 
A major failing of stochastic volatility models is that they do not describe 
the ubiquitous power law or scaling properties observed in financial data. 
These properties are observed in the volatility correlation function \cite{vol},
the PDF of high frequency price increments 
\cite{ms,tld-empirical}, 
and the temporal decay of the peak of the PDF describing the financial 
process \cite{ms}. 
These last two properties are in fact, for short time horizons, well described 
by a simple Levy process.

The growing empirical evidence of power law properties 
in financial data has generated renewed interest in the Levy paradigm. 
Recent work has shown that the problems associated with the Levy distribution 
can be simply overcome by what is known as a truncated Levy distribution (TLD).
The TLD is Levy like 
in the central part of the distribution, but has a cutoff in the far tails 
that is faster than the Levy power law tails. The cutoff will 
ensure the variance of the TLD is finite. 
Financial prices over time can
be described with the truncated Levy flight (TLF) or its continuous time limit
the truncated Levy process (TLP). The TLF is constructed from sums 
of independent and identically distributed random variables 
described by a TLD. 
Since the TLD has finite variance, the central limit theorem applies
and the TLF approachs a Gaussian distribution as desired. However what 
is interesting is the existence of a characteristic timescale separating 
the Levy and Gaussian regimes. This timescale can be arbitrarily long 
due to the stable nature of the Levy distribution. 
Mantegna and Stanley \cite{man} were the first to make 
the above observations regarding the TLF. 
They drew their conclusions based on the sums of 
independent Levy distributed random variables with a discrete cutoff 
in the tails. Inspired by these results, Koponen \cite{kop} derived an 
analytical form for the characteristic function of a TLD with an exponential 
cutoff in the tails. These theoretical results have led to several recent 
empirical studies all supporting the TLF as a simple and effective 
model of financial data \cite{ms,tld-empirical}.

The accurate modeling of financial price series is important for the 
pricing and hedging of financial derivatives such as options. 
Research on option theory with alternative pricing models has tended to focus 
on the pricing issue. It is now well known that non-Gaussian pricing models 
lead to the familiar volatility smile effect caused by the `fat' tails 
of the non-Gaussian PDF's. These effects are well known will not be the 
focus here. 
What is much less understood and discussed is the issue of option 
hedging for non-Gaussian models. This is surprising because this would seem
in many cases to be a more important issue than pricing. 
For liquid options (usually the vanilla types discussed here),
the price will be determined by the market. The real use of the model is to 
define a hedging strategy. The models option price simply provides a way 
of testing the model against the market price. 
The standard 
approach to option pricing and hedging is the Black-Scholes framework. 
What is remarkable about the Black-Scholes case is that, for Gaussian or 
log-Gaussian pricing models, there exists a 
hedging strategy which will eliminate all the risk to an option seller. 
This leads to the well known delta hedging result which says that the hedge
value is given by the derivative 
of the option price with respect to the current price of the underlying
security. 
However for more general pricing models a riskless hedge does not exist and 
in these cases the Black-Scholes framework does not tell us how to proceed. 
This is especially problematic for times close to expiry where deviations 
from Gaussian are large.
The standard approach is to simply apply the delta hedging 
procedure regardless of the pricing model used. However this approach 
is purely ad hoc as it has no clear theoretical basis.

The inadequacies of the Black-Scholes framework led Bouchaud and 
Sornette to develop a simple and more general approach to option 
theory \cite{bouchaud1,bouchaud2}. Although in general a riskless hedge does 
not exist, it is possible 
to find an optimal hedging strategy that will minimize some 
appropriate measure of risk. An obvious choice for a risk measure is 
the variance of the wealth distribution of an option seller
(or its 4th moment which would place more weight on the tails). Bouchaud 
and Sornette derived an expression for this optimal trading strategy which is 
valid for any non-Gaussian pricing model. In general the optimal trading 
strategy is not given by the delta hedge, though it is recovered for 
the special case of Gaussian or log-Gaussian models.

The outline of this paper is as follows.
In section 2 we will further test the TLP model
using high frequency data from the Australian All Ordinaries share 
market index. With the derived parameter values for the TLP, we compare 
in section 3.1 the Bouchaud-Sornette optimal hedging strategy, the TLP delta 
hedging strategy and the Gaussian delta hedging strategy. 
Although the TLP has finite variance, in the early Levy dominated 
regime a tail distribution based method to 
find an optimal trading strategy may be preferable to a moment based method. 
With this in mind, in section 3.2 we adapt to the TLP a 
simple tail distribution 
based method used by Bouchaud {\it et-al} \cite{levy-options,bouchaud2} to find the 
optimal hedging strategy for the Levy process.
Finally in section 4 we will derive a natural generalization of 
the Black-Scholes option pricing formula for the case when the underlying 
security is modeled by a geometric TLP. This demonstrates that, 
unlike for the plain Levy process, the Black-Scholes 
framework is easily adapted to the truncated Levy paradigm.

\section{The Truncated Levy Distribution}
In this section we will outline the essential properties of the TLD and the 
TLF. 
Consider a general probability density function (PDF) $P(x)$ and its 
characteristic function (CF) $\hat{P}(k)$ defined by
\begin{equation}
P(x)=\frac{1}{2\pi}\int_{-\infty}^{\infty}dk~\hat{P}(k)e^{ikx},\;\;\;
\hat{P}(k)=\int_{-\infty}^{\infty}dx~P(x)e^{-ikx}=\langle e^{-ikx}\rangle.
\end{equation} 
Moments of the distribution can be found from the CF by using 
\begin{equation}
\langle x^m(t)\rangle=
\left(i^m\frac{\partial^m}{\partial k^m}\hat{P}(k,t)\right)\bigg|_{k=0}.
\end{equation}
Normalization of the PDF requires $\hat{P}(0)=1$.
We will deal purely with symmetric $P(x)$ which in turn requires 
$\hat{P}(k)$ to be real and symmetric. 

The symmetric Levy distribution is defined by the CF \cite{lev}
\begin{equation}
\hat{L}(k)=\exp (-c^{\a}|k|^{\alpha}),\;\;\; 0<\alpha \le 2
\end{equation}
where $c$ is the scale factor and $\alpha$ is the characteristic 
exponent. 
The full PDF for the Levy distribution is only known analytically 
when $\a=1$ (Cauchy distribution) and $\a=2$ (Gaussian distribution). 
However the value of the Levy distribution is known at 
the origin where 
\begin{equation}
L(x=0)=\frac{\Gamma(1/\alpha)}{\pi\a c}
\end{equation}
and in the tails where (for $\alpha < 2$)
\begin{equation}
L(x)\rightarrow \frac{ c^{\a}\Gamma(1+\alpha)\sin \pi\alpha/2}
{\pi |x|^{\alpha+1}}, \;\;\; x\rightarrow \infty.
\end{equation}
These `fat' power law tails mean the fractional moments $\langle |x|^\mu\rangle$ 
are finite only for $\mu < \alpha$. In particular, for $\a<2$ the variance 
is infinite. 

The TLD is a generic description for a Levy distribution that has 
some cutoff far in the power law tails.
Such a cutoff will ensure that the variance of the distribution is 
finite. One possible cutoff is the exponential function 
for which the CF has been shown to be \cite{kop}
\begin{equation}
\hat{T}(k)=\exp\left[-\frac{c^{\a}}{\cos (\pi\alpha/2)}
\left((k^2+\lambda^2)^{\a/2}
\cos \left\{\a\arctan (k/\lambda)\right\}-\lambda^{\a}\right)\right],\;\;\;
\a\ne 1.
\end{equation}
With this CF the TLD in the tails takes the form
\begin{equation}
T(x)\rightarrow \frac{c^{\a}\Gamma(1+\a)\sin(\pi\a/2) e^{-\lambda|x|}}
{\pi |x|^{1+\alpha}},\;\;\;|x|\rightarrow\infty.
\end{equation}
Clearly the TLD reduces to the Levy distribution when $\lambda=0$. 
The asymmetric generalization of (2.6) was also derived in \cite{kop}. 
However in the 
paper we will deal only with the symmetric version.
Applying (2.2) to (2.6) we find the variance and kurtosis of the TLD to be  
\begin{equation}
\sigma^2=\frac{\a(1-\a)}{\cos (\pi\a/2)}c^{\a}\lambda^{\a-2}
,\;\;\;k=\frac{\cos (\pi\a/2)(\a-2)(\a-3)}{\a(1-\a)c^{\a}\lambda^{\a}}
\end{equation}
where the variance and kurtosis are defined by 
\begin{equation}
\sigma^2=\langle x^2\rangle,\;\;\;k=\frac{\langle x^4\rangle}{\langle x^2\rangle^2}-3.
\end{equation}
We can always set $c=1$ by scaling $x$ as $x\rightarrow\gamma x$. 
The CF of the scaled $x$ is again (2.6) but now with 
$c\rightarrow c\gamma$ and $\lambda\rightarrow \lambda/\gamma$. 
The kurtosis (2.8) and the exponent $\a$ will both remain fixed under scaling. 

An important consequence for option pricing is that the exponential cutoff  
in (2.7) will ensure that exponential moments $\langle e^{nx}\rangle$ 
($n$ is any real number) exist for $n\le \lambda$. 
The exponential moments 
can be found from the CF (2.6) simply by substituting 
$k^2\rightarrow -n^2$.
This is equivalent to putting $k\rightarrow in$ in (2.1) when the CF is
symmetric and can be written as a function of $k^2$. After the substitution
we can expand (2.6) in powers of $n^2/\lambda^2$ and find to first order that
\begin{equation}
\langle e^{n x}\rangle
\simeq \exp\left[\frac{n^2}{2}\langle x^2\rangle\right],
\;\;\;\lambda^2\gg n^2.
\end{equation}
In the Gaussian case ($\lambda=0,\;\a=2$) this is exact.

\subsection{Convergence to Gaussian}
Let us now consider $x$ to be the sum of $N$ independent and identically 
distributed random variables $x_i$ with a TLD defined by the CF (2.6). 
Following Mantegna and Stanley \cite{man} we can refer to $x$ as a TLF.
The CF of $x$ will be
\begin{equation}
\hat{T}(k,N)=\exp\left[-\frac{c^{\a}N}{\cos (\pi\alpha/2)}
\left((k^2+\lambda^2)^{\a/2}
\cos \left\{\a\arctan (k/\lambda)\right\}-\lambda^{\a}\right)\right],\;\;\;
\a\ne 1
\end{equation}
which is the same as (2.6) but with $c^{\a}\rightarrow 
Nc^{\a}$. First consider the special case when each $x_i$ has a PDF described by
a Levy distribution ($\lambda=0$). The Levy distribution is also known as 
the Levy {\it stable} distribution. This is because the PDF of the rescaled 
variable $x N^{-1/\a}$ is the same as that of $x_i$, that is the Levy 
distribution is stable under addition (the central limit theorem does not 
apply since the Levy distribution has infinite variance). 
We can also refer to this as the 
{\it scaling} or {\it fractal} property of the Levy distribution.
On the other hand, for $\lambda>0$, the variance 
of $x_i$ is finite so the PDF of $x$ must approach a Gaussian by the 
central limit theorem. What is remarkable is that, as first pointed out by 
Mantegna and Stanley \cite{man}, the convergence of $x$ to a Gaussian occurs 
extremely slowly due to the stable property of the Levy distribution. 

We can derive the crossover time $N_c$ for $x$ to converge to a Gaussian 
as follows. From (2.8) we 
know that the variance and kurtosis of $x$ are given by
\begin{equation}
\sigma^2(N)=\frac{\a(1-\a)}{\cos(\pi\a/2)}c^{\a}\lambda^{\a-2}N,\;\;\;
k(N)=\frac{\cos(\pi\a/2)(\a-2)(\a-3)}{\a(1-\a) c^{\a}\lambda^{\a}N}.
\end{equation}
A useful qualitative model of the distribution of $x$ is that of a Gaussian
in the central part out to the scale determined by the square root of the 
variance (2.12) (and therefore growing as $\sqrt{N}$). 
Beyond this scale we can think of the distribution being 
described by the tails (2.7) (with $c^{\a}\rightarrow Nc^{\a}$) 
which are slowing being consumed by the 
Gaussian part.  
We can then ask at what time $N_c$ does the scale set by the square root of 
the variance equal the cutoff scale $\lambda^{-1}$. We easily find
the crossover time $N_c$ to be
\begin{equation}
N_c=\frac{\cos(\pi\a/2)}{\a(1-\a)}c^{-\a}\lambda^{-\a}.
\end{equation}
A more rigorous method to derive this timescale is to find the time 
for the kurtosis of $x$ to decay away. From (2.12) we see that this timescale 
is consistent with (2.13) which justifies the simple qualitative picture 
of the convergence to a Gaussian. We will find this picture helpful later when 
option hedging is considered. Mantegna and Stanley \cite{man} first 
derived the timescale $N_c\sim c^{-\a}\lambda^{-\a}$ using a method based on 
the probability of $x$ returning to the origin (this method gives a slightly 
different $\alpha$ dependent coefficient). This result can also be obtained
from the Berry-Esseen theorem \cite{shl} and will be independent of the 
precise form of the cutoff. Clearly $N_c$ can be as large as one wishes by 
making $\lambda$ small enough.

\subsection{Parameter Fitting}
Parameter fitting to a TLF is no more complex than fitting data to a Levy 
distribution. The parameters $\a$ and $c$ can be obtained by fitting a Levy 
distribution to price increments (or log increments) on the smallest timescale 
such that linear correlations are negligible. A good rule of thumb is 30 minute
increments. The timescale needs to be small so that we are operating well in 
the Levy regime of the TLF.
Once $c$ and $\a$ are known we can extract 
the cutoff parameter $\lambda$ from the variance of the dataset 
using (2.12). This will fit the variance of the data for a 
particular time $N$. From the point of view of option pricing fitting to 
the daily variance would seem the most appropriate choice. 
It would be appropriate to fit $\lambda$ using a large dataset to ensure a good 
description of the tails. Once $\a$ and $\lambda$ are determined 
we can expect these parameters to be relatively stable despite the well 
documented non-stationary behavior associated with a stochastic variance.
These effects can be accounted for by a time dependent scale factor $c$.
The non-stationary behavior is associated with variance measurements 
over small data sets (typically 1 month). This data will only sample the 
central part of the PDF which compared to $c$ is relatively insensitive to 
$\lambda$. The relative stability of the exponent $\a$ over time might be expected since
it is invariant to scale changes. As a consequence sums of Levy distributed 
random variables with the same exponent $\a$, but different scale factor $c$, 
will still have a distribution described by the exponent $\a$.
This stability has been confirmed in empirical studies on the S\&P 500 
index \cite{ms}.

There are several methods for fitting data to a Levy distribution (see 
Rachev and Mittnik \cite{levy-empirical} for a review). 
Here we adopt the method used by Mantegna and Stanley \cite{ms}. 
It is based on the 
relation
\begin{equation}
T(x=0,N)\simeq L(x=0,N)=\frac{\Gamma(1/\alpha)}{\pi\a c N^{1/\a}}, \;\;\;
N\ll N_c
\end{equation}
which will be a good approximation in the Levy regime of the TLF ($N\ll N_c$). 
In this method we can 
extract $c$ and $\a$ by fitting a power law decay to the probability 
of return to the origin for times $N\ll N_c$. 
Alternatively we can take the log of (2.14) and extract 
$c$ and $\a$ by a straight line fit. This is a very practical method because 
it is simple, it can be used on relatively small datasets (since $T(0,N)$ is the 
maximum of the distribution), and because of the easy availability 
of high frequency financial data. It is somewhat different to other methods in 
that it fits the peak of the distribution at various times rather than  
fitting the whole distribution at a single time.

We will demonstrate the parameter fitting process using high 
frequency equity data. We have used the All Ordinaries Index (AOI) which is the 
major Australian equity index. The dataset comprised the value of the AOI 
at 5 minute intervals from 1993-May/1997. From this, datasets describing the 
raw change in the AOI were constructed for 30 minutes, 1 hour, 2 hours, 3 hours 
and 1 day. The mean was then subtracted from these datasets. The  
time of 30 minutes was considered to be a good minimum time 
for which changes in the AOI could be considered to have negligible 
linear correlations. This is necessary in order to justify the use of (2.14). 
From these datasets the probability of zero change in the AOI was found 
and plotted against time in figure 1 (time is in units of 30 minutes 
with 1 trading day equal to 6 hours or $N=12$).

The data of figure 1 was very well fitted 
to a power law decay curve. From the curve of best fit and (2.14) it was found 
that $\a\simeq 1.2$ and $c\simeq 1.1$. For comparison in figure 1 the 
fit of $P(0)$ to a Gaussian process is shown where the Gaussian is fitted to 
the daily variance of 226 (as would occur in practice from an option 
pricing perspective) and extrapolated back to 30 minutes. 
In figure 2 the empirical PDF for the 30 minute data is shown along with 
the Levy PDF defined by the CF (2.11) with the parameters $\a=1.2, c=1.1,\lambda=0$ and $N=1$. 
We also show the Gaussian PDF which has been 
fitted to the variance of the daily data and extrapolated back to 30 minutes. 
The Levy PDF can be seen to be in 
good aggreement with the empirical PDF all the way out to  
to $\pm 50$ points which is approximately 12 standard deviations. 
The tails of the distribution are quite sensitive to the exponent $\a$ 
which we demonstrate by also showing the Levy PDF for $\a=1.4$. 
From this we can see that simple fitting method used has been quite 
effective.
These results along with others \cite{levy-empirical,ms,tld-empirical} 
further support the LF/TLF model of financial data.

To fit $\lambda$ the dataset used was that of daily changes in the AOI between 
1987-96. A large data set is required to fit $\lambda$ since this relates 
to the tails of the distribution. Substituting the daily variance of 345 and 
$N=12$ into (2.12), and using the same values for $\a$ and $c$ derived 
previously, we find that $\lambda=1/80$. 
With these parameter values 
we find that the Levy-Gaussian crossover time (2.13) is $N_c\simeq 222$, or 
approximately 19 trading days. This is very close to 1 trading month (21 days)
which previous studies have shown to be the timescale for which the Gaussian 
description becomes accurate \cite{ak,ms}.
Since $N_c\gg 1$ day the use of (2.14) out 
to 1 trading day is justified since this is well within the Levy regime of 
the TLF.

The daily dataset from 1987-96 is quite symmetric except in the far tails. 
The largest positive deviation in the dataset
is 80 points. On the other hand there are several negative deviations beyond 
100 points which all occured around the 1987 crash of -520 points.  
These facts suggest that an appropriate asymmetric 
model would be a symmetric Levy distribution with asymmetric cutoff parameters,
rather than an asymmetric Levy distribution with symmetric cutoff 
parameter. The later model \cite{kop} would give asymmetry in the whole distribution 
rather than just in the far tails. In option pricing generally only part 
of the probability distribution is important. This suggests that
a simple and effective approach to asymmetry would be to choose a 
cutoff parameter that best fits the part of the distribution that is 
relevant. To illustrate this we have derived the cutoff parameter 
$\lambda_+=1/41$ from 
the variance of positive deviations (203), and 
the cutoff parameter $\lambda_-=1/122$ derived from the 
variance of negative deviations (487). This latter cutoff parameter leads 
to a crossover time of approximately 30 days. 
We have used $\a=1.2, c=1.1$ and  $N=12$
as before.

\section {Optimal Option Hedging}
In this section we will be concerned with the optimal hedging of 
call options for times to expiry which are less than the Levy-Gaussian 
crossover time. With 1 trading month as a typical crossover timescale we would 
be interested in times to expiry roughly less than 10 trading days. In this 
case the drift and probability of negative prices are negligible and 
we can consider the TLP as our model for the financial data. 

Consider first the price $C(S_0,E,t)$ of a European call option at current 
time $t=0$, with exercise price $E$ due to expire in a time $t$. 
When the time to expiry is small the returns and interest rates can be 
neglected. The option price is then very well approximated by
\begin{equation}
C(S_0,E,t)\simeq \Bigl\langle {\rm max}(S-E,0)\Bigr\rangle
=\int^{\infty}_EdS~(S-E)P(S,t|S_0,0)
\end{equation}
where $P(S,t|S_0,0)$ is the driftless PDF of the underlying asset price $S(t)$. 
We know in advance that (3.1) will 
lead to the familiar volatility smile effect caused by the `fat' tails 
of non-Gaussian PDF's like the TLD. 
These effects are well known and we will not consider them
further. 
What is much less understood and discussed is the issue of option 
hedging for non-Gaussian models. It is to this that we now turn. 

\subsection{Variance based hedging strategy}
A framework which can address the hedging issue is the 
Bouchaud-Sornette approach to option pricing and hedging 
\cite{bouchaud1,bouchaud2}.
This approach starts by finding 
the variation in wealth for a call option seller. 
They find the wealth variation 
between times $0$ and $t$ can be written as
\begin{equation}
\Delta W|_0^t= C(S_0,E,t)-{\rm max}(S(t)-E,0) +\int^t_0d\tau~\phi(S_\tau)
\dot{S}(\tau)
\end{equation}
where the first term is the option premium received at $t=0$, the second term 
describes the payoff at expiry $t$ and the third term describes the effect 
of trading where $\phi(S_{\tau})$ ($S_\tau=S(\tau)$) is the amount of stock 
held.
The interest rate is set at zero for clarity which will be a good 
approximation for short-medium term options.  The option price is found 
by requiring $\langle\Delta W|_0^t\rangle=0$. This leads to 
\begin{equation}
C(S_0,E,t)=\Bigl\langle {\rm max}(S(t)-E,0)\Bigr\rangle -\int^t_0d\tau~
\langle\phi(S_\tau)\rangle
\langle\dot{S}(\tau)\rangle
\end{equation}
where the increment $\dot{S}(\tau)$ is in the future and 
assumed to be independent of $S(\tau)$ (this assumption can be relaxed 
\cite{bouchaud1}). 
What is remarkable about the Black-Scholes case (Gaussian or log-Gaussian 
models) is that there exists a trading strategy $\phi^*(S_\tau)$ such that 
$\Delta W|_0^t$ can be made to vanish. In this case the rate of return 
dependence in (3.3) cancels and the Black-Scholes 
result follows \cite{bouchaud1}. 
For more general pricing models a riskless 
hedge does not exist and in these cases the Black-Scholes framework fails. 
However the Bouchaud-Sornette method is easily adapted to these more realistic 
situations. Although in general a riskless hedge does not exist, it is possible 
to find an optimal trading strategy $\phi^*(S_{\tau})$ that will minimize some 
appropriate measure of risk. An obvious choice for a risk measure is 
the variance of the wealth distribution $\langle\Delta W^2|_0^t[\phi]\rangle$,
or its 4th moment which would place more weight on the tails. 
The optimal trading strategy can be easily computed for any PDF describing 
the price $S(t)$. In these more 
general cases the option price (3.3) (evaluated at $\phi^*$) will depend on the 
rate of return. However the return can be safely set to zero for options less 
than a few months to expiry (see Aurell {\it et-al} \cite{tld-empirical}).
In this case the second term on the right hand side of (3.3) will vanish 
and (3.1) will be a good approximation.
However the option price (3.3) will need to be corrected 
by a risk premium whose scale will be set by the residual risk
$\sqrt{\langle\Delta W^2|_0^t[\phi^*]\rangle}$. This 
risk premium can account for the bid-ask spreads in option prices. 

Consider the special case of a pricing model in which the price increments are 
uncorrelated, stationary and have zero mean (these assumptions can be relaxed).
This includes the TLP and will be a good model for options roughly less than 
a month to expiry. In this case the hedging strategy that minimizes the 
variance of the wealth distribution has been shown to be 
\cite{bouchaud1,bouchaud2}
\begin{equation}
\phi^*(S_0,E,t)=\frac{1}{\sigma^2 t}\int^{\infty}_E dS~(S-E)(S-S_0)
P(S-S_0,t),
\end{equation}
where $P(S-S_0,t)$ is the conditional PDF of the pricing model with 
initial price $S_0$ and $\sigma^2 t$ is the variance of the model. 
The optimal hedge $\phi^*$ ranges between 0 and 1 since we are dealing 
with an option on one unit of the underlying security.
For at-the-money 
options ($E=S_0$) we easily find from (3.4) that $\phi^*=1/2$ for any pricing 
model. For options
out-of-the-money ($E>S_0$) we have $\phi^*<1/2$ and for options 
in-the-money ($E<S_0$) we have $\phi^*>1/2$. 

For the special case of the Gaussian pricing model with PDF
\begin{equation}
G(S-S_0,t)=\frac{1}{\sqrt{2\pi\sigma^2
t}}\exp\left(-\frac{(S-S_0)^2}{2\sigma^2 t}\right),
\end{equation}
the optimal hedging strategy (3.4) reduces to the Gaussian Black-Scholes result 
\begin{equation}
\phi^*(S_0,E,t)=\frac{\partial}{\partial S_0}\int^{\infty}_E dS~(S-E)
G(S-S_0,t)=N\left(\frac{S_0-E}{\sigma\sqrt{t}}\right)
\end{equation}
where $N$ is defined in (4.17). This result can also be shown to be the 
probability of exercise of the option. 
This hedging strategy is clearly just the 
derivative of the option price with respect to $S_0$. This is the well 
known delta hedging result and can be expressed more generally as 
\begin{equation}
\Delta(S_0,E,t)=\frac{\partial}{\partial S_0}C(S_0,E,t)
\simeq\frac{\partial}{\partial S_0}\int^{\infty}_E dS~(S-E)P(S-S_0,t)
\end{equation}
where we have used the approximate option price (3.1). This is a riskless 
hedging strategy only for Gaussian and log-Gaussian models. 
However it can be used to give the hedge value for other pricing models.

Consider the PDF for the TLP which has the tail form
\begin{equation}
T(S-S_0,t)\simeq \frac{ c^{\a}t\Gamma(1+\alpha)\sin \pi\alpha/2}
{\pi |S-S_0|^{1+\alpha}}\exp(-\lambda|S-S_0|), \;\;\; 
\frac{|S-S_0|}{\sqrt{\sigma^2 t}} >1.
\end{equation}
The scale where this approximation becomes valid is set by the Levy process
diffusion scale $ct^{1/\a}$. The variance of the TLP will always be 
greater than this scale for times less than the Levy-Gaussian crossover 
time. We will always be interested in times less than this which 
makes the condition on $|S-S_0|$ appropriate. We wish to compare for this
model, the optimal Bouchaud-Sornette hedging strategy (3.4) with the 
delta hedge (3.7). We will also compare the TLP hedges with the 
Gaussian delta hedge (3.6).
Applying the PDF (3.8) to (3.4) we find that the optimal Bouchaud-Sornette 
hedge becomes
\begin{equation}
\phi^*(S_0,E,t)\simeq \frac{\Gamma(1+\a)\sin\pi\a}{2\pi\a (\a-1)^2}
\left[y^{2-\a}e^{-y}+(1-y-\a)\Gamma(2-\a,y)\right], 
\;\;\;\frac{E-S_0}{\sqrt{\sigma^2 t}} > 1,
\end{equation}
where
\begin{equation}
y=\left(\frac{t}{t_c}\right)^{1/2}\frac{|E-S_0|}{\sqrt{\sigma^2 t}}=
\lambda|E-S_0|
\end{equation}
and
\begin{equation}
\Gamma(a,x)=\int^{\infty}_xdt~e^{-t}t^{a-1}
\end{equation}
is the incomplete gamma function. 
In (3.10), $t_c$ is the same as the discrete time Levy-Gaussian crossover 
time defined in (2.13).
For $S_0>E$ we have
\begin{equation}
\phi^*(S_0,E,t)\simeq 1-\frac{\Gamma(1+\a)\sin\pi\a}{2\pi\a (\a-1)^2}
\left[y^{2-\a}e^{-y}+(1-y-\a)\Gamma(2-\a,y)\right], 
\;\;\;\frac{S_0-E}{\sqrt{\sigma^2 t}} > 1.
\end{equation} 

Consider next the delta hedging strategy based on the TLP. Applying the PDF
(3.8) to (3.7) we find
\begin{equation}
\Delta(S_0,E,t)\simeq\frac{t}{t_c}
\frac{\Gamma(1+\a)\sin\pi\a}{2\pi\a^2 (\a-1)^2}
\left[y^{1-\a}e^{-y}-(\a-1)y^{-\a}e^{-y}-\Gamma(2-\a,y)\right], 
\;\;\;\frac{E-S_0}{\sqrt{\sigma^2 t}}> 1,
\end{equation}
and for $S_0>E$
\begin{equation}
\Delta(S_0,E,t)\simeq 1-\frac{t}{t_c}
\frac{\Gamma(1+\a)\sin\pi\a}{2\pi\a^2 (\a-1)^2}
\left[y^{1-\a}e^{-y}-(\a-1)y^{-\a}e^{-y}-\Gamma(2-\a,y)\right], 
\;\;\;\frac{S_0-E}{\sqrt{\sigma^2 t}}> 1.
\end{equation}
The Gaussian delta hedge (3.6) is exactly equal to the probability 
of exercise of the option for the Gaussian pricing model (3.5). 
The TLP delta hedge (3.13-14) is 
also equal to the probability of exercise when calculated with the tail 
approximation (3.8).

In figures 3 and 4 we plot and compare the TLP Bouchaud-Sornette optimal 
hedge (3.9), the TLP delta hedge (3.13) and the Gaussian delta hedge (3.6). 
We are 
using $E>S_0$ which means the options are out-of-the-money and have a low
hedge value and probability of exercise. We plot the hedge value against 
the ratio of $E-S_0$ and the standard deviation of the pricing model. 
No parameter values need to be specified in the Gaussian delta hedge 
(3.6). For the hedges (3.9) and (3.13), we know from (3.10) that
we need to specify the exponent $\a$ and the ratio of the time to expiry $t$ 
and the Levy-Gaussian crossover time $t_c$. 
In section 2 we obtained $\a=1.2$ and a Levy-Gaussian 
crossover time of approximately 20 trading days. We have used these parameters 
with the expiry times of 1 and 5 days in figures 3 and 4 respectively. 
In figure 3 we see that Bouchaud-Sornette optimal hedge leads to hedge values 
that are, in percentage terms, significantly greater than those obtained with 
both delta hedges. 
The relative difference increases as $E-S_0$ grows. In figure 5 the 
differences are not as great, but in percentage terms they are still very 
large.

\subsection{Distribution based hedging strategy} 
The Bouchaud-Sornette method will fail for models such as the Levy process 
which have infinite
moments. Although the TLP has finite variance, in the early Levy dominated 
regime a distribution based method to 
find an optimal trading strategy may be preferable to a moment based method. 
Below we will adapt to a truncated Levy model a simple method used by 
Bouchaud {\it et-al} \cite{levy-options,bouchaud2} to find the optimal hedging 
strategy for Levy models.

Consider the case where $\phi$ is the final hedge value some time
$t$ from expiry. 
Then from (3.2) we find that the change in wealth over this period is (ignoring the premium)
\begin{equation}
\Delta W = I\Bigl((\phi-1)(S_t-S_0) + E-S_0\Bigr) +(1-I)\phi (S_t-S_0),\;\;\;
0\le\phi\le 1
\end{equation}
where  $I=1$ for $S_t>E$ and $I=0$ for $S_t< E$. Now we make the key assumption 
that $I$ can be treated as an independent random variable that takes on 
values 1 and 0 with probabilities ${\cal P}$ and $1-{\cal P}$,
where ${\cal P}$ is the probability the option will be exercised. 
Note that both $\phi$ and ${\cal P}$ are functions of $(S_0,E,t)$. 
We can write the change in wealth 
$\Delta W|_{S_t-S_0}$, due only to the change in price as 
\begin{equation}
\Delta W|_{S_t-S_0}=I(1-\phi)(S_0-S_t)+(1-I)\phi(S_t-S_0).
\end{equation}
In most realizations of $I$ and $S_t-S_0$, this will lead to a loss which we 
denote
by $l$. We assume that $S_t-S_0$ is described by a TLP which has the PDF
(3.8) in the tails. 
Due to the independence assumption of $I$ we can easily write down the tail 
PDF of the loss $l$ using (3.8). From this we find that the probability of a 
loss greater than $l_*$ is 
\begin{equation}
{\cal P}(l> l_*)\sim \Bigl(1-{\cal P}\Bigr)\phi^{\a}
\int^{\infty}_{l_*}dl~\frac{e^{-\lambda_- l/\phi}}{l^{1+\a}}
+{\cal P}(1-\phi)^{\a}\int^{\infty}_{l_*}dl~
\frac{e^{-\lambda_+ l/(1-\phi)}}{l^{1+\a}}.
\end{equation}
We can write $l_*=1/(2\lambda_*)$ which is the average loss suffered to an event 
$\pm 1/\lambda_*$. Thus we can consider minimizing losses above
$l_*$ to be equivalent to minimizing losses to events beyond $\pm 1/\lambda_*$. 
So after performing the integrals we can write (3.17) as
\begin{equation}
{\cal P}(l> l_*)\sim \Bigl(1-{\cal
P}(S_0,t)\Bigr)\lambda_-^{\a}\Gamma(-\a,\beta_-)
+{\cal P}(S_0,t)\lambda_+^{\a}\Gamma(-\a,\beta_+)
\end{equation}
where 
\begin{equation}
\beta_-=\frac{\lambda_-}{2\lambda^*\phi},\;\;\;\beta_+=
\frac{\lambda_+}{2\lambda^*(1-\phi)}.
\end{equation}

We wish to find the optimal trading strategy $\phi^*$ that will 
minimize the probability (3.18). This is obtained by 
by taking the derivative of (3.18) with respect to $\phi$, setting the left 
hand side to zero to obtain
\begin{equation}
0=\Bigl(1-{\cal P}\Bigr)\phi^{\a-1}\exp(-\beta_-)-{\cal P}
(1-\phi)^{\a-1}\exp(-\beta_+),
\end{equation}
and solving this equation for $\phi^*$ as a function of ${\cal P}$.
For $\lambda=0$, (3.20) becomes independent of $\lambda_*$ and it can be solved 
to obtain the Levy optimal hedge
\begin{equation}
\phi^*(S_0,E,t)=\frac{{\cal P}^{\xi}(S_0,E,t)}{{\cal P}^{\xi}(S_0,E,t)
+(1-{\cal P}(S_0,E,t))^{\xi}},\;\;\;\xi=1/(\a-1)
\end{equation}
first derived by Bouchaud {\it et-al} \cite{levy-options,bouchaud2}. 
For $\a=2$ the Gaussian delta hedge $\phi^*={\cal P}$ is recovered. 

The Levy hedge (3.21) is of fundamental interest as it shows that a hedging 
strategy can be derived even for a model with infinite variance. 
However what is
required in practice is a trading strategy that interpolates between the Levy 
hedge and the 
Gaussian hedge over the crossover timescale (2.13). 
This follows from the central limit theorem which ensures that the optimal 
hedge tends to the Gaussian hedge (this is not 
the case for the geometric TLP discussed in section 4).
Clearly we want to choose $\lambda_*^{-1}$ to be the smallest possible value 
such that the tail form of the PDF will hold. The scale beyond which this holds 
is set by the Levy diffusion scale $ct^{1/\a}$.
A good choice is  
$\lambda_*^{-1}=\sigma\sqrt{t}$ which is the standard deviation of 
$S_t-S_0$. This scale will always be greater than the Levy diffusion scale for 
times less than the crossover timescale. 
This is fine since this method of hedging will only be appropriate for 
these times. The choice for $\lambda_*$ is also consistent with the qualitative
picture, discussed in section 2.1, for the convergence of the TLP to a
Gaussian.

In figure 5 we plot the optimal hedging strategy against the probability of exercise 
obtained by a numerical solution of (3.20). We have chosen as parameters those obtained 
in section 2.2 ($\a=1.2$ and $\lambda=1/80$) which gave a Levy-Gaussian crossover time of 
approximately 19 trading days. Using $\lambda_*^{-1}=\sigma\sqrt{t}$, we plot 
the optimal hedge at 2 days ($\lambda/\lambda_*\simeq0.33$), 10 days 
($\lambda/\lambda_*\simeq 0.73$)
and 20 days ($\lambda/\lambda_*\simeq 1.03$) along with the Gaussian 
hedge ($\a=2,~\lambda=0$) and the Levy hedge ($\lambda=0$). 
The hedging strategies have the very nice feature of a well defined evolution 
from the Levy hedge (3.21) to the Gaussian hedge in about 10 days. 
Between 10 and 20 days there 
is a much slower creep away from the Gaussian hedge. From this we can
tentatively conclude that
the hedging strategy defined by (3.20) could be reliably used out to time 
$t_*$ defined by 
$\lambda_*\simeq \lambda/0.73$, or approximately $1/2$ the crossover time 
(2.13). Beyond this the Bouchaud-Sornette variance based method
would be appropriate. In figure 5 we see that for options with a small 
probability of exericise,
the difference between the optimal hedge and the Gaussian Black-Scholes 
hedge is large in percentage terms. This is consistent with what was found in
section 3.1. 

Figure 6 is similar to figure 5 except that we have used the asymmetric 
cutoff parameters $\lambda_+=1/41$ and $\lambda_-=1/122$ discussed in section 2.2. 
The symmetric cutoff parameter $\lambda=1/80$ was used in the variance required to describe
the growth in $\lambda_*$. This case gives an optimal hedge that involves holding less 
stock than in the symmetric case. This is what we expect since the asymmetric case has more 
weight in the negative tails and negative price changes favor holding zero 
stock. Plots of the ratio of the asymmetric/symmetric hedging strategies of 
figures 5 and 6 approximately range between 0.7 and 1 for exercise
probabilities between 0 and 1 respectively.

\section{Option Pricing}
Here we will derive a generalization of the Black-Scholes 
option pricing formula for the case where the underlying security is modeled 
by a geometric TLP. We will see that the Black-Scholes 
framework is easily adapted to the truncated Levy paradigm, a feature not 
shared by the plain Levy or geometric Levy process.

The Black-Scholes option pricing theory \cite{bs} is based on the 
geometric Brownian Motion (GBM) paradigm of financial market dynamics 
\cite{hull}. 
A natural generalization 
of GBM is to define the geometric TLP $S(t)$ by  
\begin{equation}
S(t)=S_0\exp\left[x(t)+\mu t -\frac{1}{2}\sigma^2 t\right],
\end{equation}
where $x(t)$ is a stochastic process defined by the CF
\begin{equation}
\hat{T}_g(k,t)=\exp\left[-\frac{c_x^{\a}t}{\cos (\pi\alpha/2)}
\left((k^2+\lambda_x^2)^{\a/2}
\cos \left\{\a\arctan (k/\lambda_x)\right\}-\lambda_x^{\a}\right)\right],\;\;\;
\a\ne 1,
\end{equation}
$\sigma^2 t$ is the variance of $x(t)$ ($\sigma^2$ is defined in (2.8)) 
and $\mu$ will be shown to be the rate of return. 
A time dependent 
variance can be considered by replacing $\sigma^2 t$ with 
$\int^t_0ds~\sigma^2(s)ds$. This time dependence will
derive from a time dependent scale factor $c_x(t)$ in (4.2) with $\a$ and
$\lambda_x$ kept constant. GBM is 
recovered by setting $\a=2$ and $\lambda_x=0$ in the CF (4.2).
Using (2.8) this gives $2c_x^2=\sigma^2$.

Using (2.10) we find the moments of the geometric TLP are
\begin{equation}
\langle S^n(t) \rangle
\simeq S_0^n\exp\left[n\mu t+n(n-1)
\sigma^{2} t/2\right], \;\;\;\lambda_x^2 \gg n^2.
\end{equation}
From this we see that $\mu$ is the rate of return. This expression is an approximation 
to first order in $n^2/\lambda_x^2$ and is exact for GBM. The approximation 
breaks down 
for very high order moments which is direct consequence of the absence of a central limit 
theorem for multiplicative processes like (4.1).
But how do we know that the condition $\lambda_x^2 \gg n^2$ will hold?
Consider short timescales where $x(t)$ is small and we can write (4.1) as
\begin{equation}
S(t)\simeq S_0 + S_0 x(t).
\end{equation}
Under this approximation $S(t)$ reduces to the plain TLP.
Equation (4.4) will be a valid approximation when $\sqrt{\sigma^2 t}\ll 1$. 
This will apply for time periods typically of the order of 
1 month or less.
Fitting data to raw changes in the price $S(t)$, as in section 2.2, effectively 
fits $S_0x(t)$ to a TLP which is described by the parameters $\a$, $c$ and 
$\lambda$. We can 
then extract the parameters $c_x,\lambda_x$ of the CF of $x(t)$, by the 
relations $c_x = c/S_0$ and $\lambda_x = S_0\lambda$ ($\a$ is invariant 
under scaling). 
We know that $\lambda_x$ must obey $\lambda_x\gg 1$ because the original 
cutoff scale $\lambda^{-1}$
of the TLP must be much smaller than $S_0$ to ensure the price 
stays positive. We have found that fitting daily data to change in 
log price leads to $\lambda_x$ typically in the range of 20-30.

Lets now consider the price $C(S_0,E,t)$ of a European call option at time $t=0$ with 
exercise price $E$ due to expire in time $t$. We will assume that the price 
of the underlying asset $S(t)$ follows a geometric TLP (4.1) with $S_0$ the 
current price. 
A simple but ad hoc approach to option pricing  is to apply the
the risk-neutral approach \cite{hull} in exactly the same way it is used with 
the GBM model. 
In this case the option price if given by
\begin{equation}
C(S_0,t)=e^{-rt}\Bigl\langle {\rm max}(S-E,0)\Bigr\rangle
=e^{-rt}\int^{\infty}_EdS~(S-E)T_g(S,t|S_0,0)
\end{equation}
where $T_g(S,t|S_0,0)$ is the PDF of the geometric TLP with $\mu$ set equal 
to $r$.
Using $\int^{\infty}_E\rightarrow \int^{\infty}_0-\int^{E}_0$, 
the normalization of $T_g(S,t|S_0,0)$ and (4.3)
we find that
\begin{equation}
C(S_0,t)\simeq S_0-Ee^{-rt}-e^{-rt}\int^{E}_0dS~(S-E)T_g(S,t|S_0,0),\;\;\;
\lambda_x^2\gg 1.
\end{equation}
We can write the PDF of $S(t)$ in the form
\begin{equation}
T_g(S,t|S_0,0)dS=\frac{dx}{\pi}\int_{0}^{\infty}dk~\hat{T}_g(k,t)\cos kx,
\end{equation}
where $\hat{T}_g(k,t)$ is the CF of $x(t)$ defined by (4.2), and from (4.1) 
we have (with the risk-neutral measure)
\begin{equation}
x=\ln(S/S_0)-rt+\frac{1}{2}\sigma^2 t.
\end{equation}
We can substitute (4.7) into (4.6) and change the order of integration. 
Using the integrals 
\begin{equation}
\int^{x_e}_{-\infty}dx~\cos kx=\frac{\sin kx_e}{k}+\pi\delta(k)
\end{equation}
and
\begin{equation}
\int^{x_e}_{-\infty}dx~e^x\cos kx=e^{x_e}\frac{(\cos kx_e+k\sin kx_e)}{1+k^2}
\end{equation}
we find the call option price is given by 
\begin{equation}
C(S_0,E,t)\simeq S_0-\frac{1}{2}Ee^{-rt}+\frac{Ee^{-rt}}{\pi}
\int^{\infty}_0dk~\hat{T}_g(k,t)
\left(\frac{\sin kx_e-k\cos kx_e}{k(1+k^2)}\right),\;\;\;\lambda_x^2\gg 1
\end{equation}
with $x_e$ defined by
\begin{equation}
x_e=\ln(E/S_0)-rt+\frac{1}{2}\sigma^{2} t.
\end{equation}
We found $\lambda_x$ to be in the range 20-30. This means the approximation 
(4.11) will be very good.

The Black-Scholes result is recovered from (4.11) by writing
\begin{equation}
\hat{T}_g(k,t)=\exp(-\sigma^2 tk^2/2)
\end{equation}
and using the identities
\begin{equation}
\frac{\sin kx_e-k\cos kx_e}{k(1+k^2)}=\frac{\sin kx_e}{k}-
\frac{(k\sin kx_e+\cos kx_e)}{1+k^2},
\end{equation}
\begin{equation}
\int^{\infty}_0dk~\frac{e^{-pk^2}}{\gamma^2+k^2}(k\sin kx+\gamma\cos kx)=
\pi e^{p\gamma^2-x\gamma}N\left(\frac{x-2\gamma p}{\sqrt{2p}}\right),
\;\;\;\gamma >0
\end{equation}
and
\begin{equation}
\int^{\infty}_0dk~e^{-pk^2}\frac{\sin kx}{k}=
\frac{\pi}{2}-\pi N\left(\frac{-x}{\sqrt{2p}}\right)
\end{equation}
where 
\begin{equation}
N(x)=\frac{1}{\sqrt{2\pi}}\int^x_{-\infty}dy~e^{-y^2/2},\;\;\;N(x)=1-N(-x).
\end{equation}
We then obtain the Black-Scholes option pricing formula \cite{bs}
\begin{equation}
C(S_0,E,t)=S_0 N(d)
-Ee^{-rt}N(d-\sqrt{\sigma^2 t})
\end{equation}
where
\begin{equation}
d=\frac{\sigma^2t-x_e}{\sqrt{\sigma^2 t}}.
\end{equation}
Its important to emphasize that the option price (4.11) will {\it not}
approach the Black-Scholes result (4.18) as the time to expiry increases. 
This is because the central limit theorem does not apply to multiplicative 
processes like the geometric TLP \cite{mult}.

\section{Conclusion}
In this paper we have further demonstrated that the TLP can effectively model 
the 
empirical successes of both the Levy and Gaussian distributions at short and 
long timescales respectively. In option theory a major disincentive for using 
non-Gaussian 
based models is the absence of a riskless hedge. 
This makes it impossible to apply the Black-Scholes option pricing 
framework in anything other than an ad hoc way.
We therefore discussed how the Bouchaud-Sornette approach to option 
theory provides a relatively simple solution to the problem of option pricing 
and optimal hedging in non-Gaussian models. We applied this optimal hedging 
strategy to the TLP pricing model and compared it with the delta hedging 
strategy for both the TLP and Gaussian models. Significant differences 
were found. We then discussed an alternative
and computationally simple tail distribution based hedging strategy appropriate for 
the Levy regime of the TLP. 
For a TLP with exponential cutoff, the exponential 
moments exist and this allowed us to derive in section 4 a natural 
generalization of the Black-Scholes option pricing formula for the 
case of a geometric TLP pricing model. This is not possible for the 
geometric Levy process, an important reason for the lack of interest in this 
model. 

Further theoretical work could involve deriving CF's of TLD's with alternative 
cutoffs such as stretched exponential tails, or models with asymmetric cutoff 
parameters. Stretched exponential tails are interesting technically because the
Laplace transform of the distribution would then exist \cite{sor}. 
An asymmetric cutoff will give a PDF that is symmetric in the center but
asymmetric in the far tails. This was how asymmetry occurred in the
daily data set of the AOI considered in this paper. 
Also of great practical importance is the construction of efficient algorithms 
for the numerical simulation of random variables with a TLD. It may be possible
to adapt existing methods for the simulation of 
Levy distributed random variables \cite{sim}. With simulations we could test 
the effectiveness of the tail based hedging strategy against the 
variance based method. 
This is important since uncontrolled approximations had to be made in deriving 
the tail based strategy. 
The simple TLP model studied here does not address the apparent stochastic 
dynamics of the volatility along with its long range power law correlation 
function \cite{vol}. A challenge for the future is to construct pragmatic 
models which can describe these properties as well as the 
those of the distribution considered here. 

\vskip 1cm
\noindent{\bf Acknowledgement}: 
I would like to thank the Australia Research Council for their
generous support of this research
through an Australian Postdoctoral Research Fellowship.


\newpage
\begin{figure}[tbp]
\epsfxsize=14.0cm
\epsfysize=10.0cm
\centering{\ \epsfbox{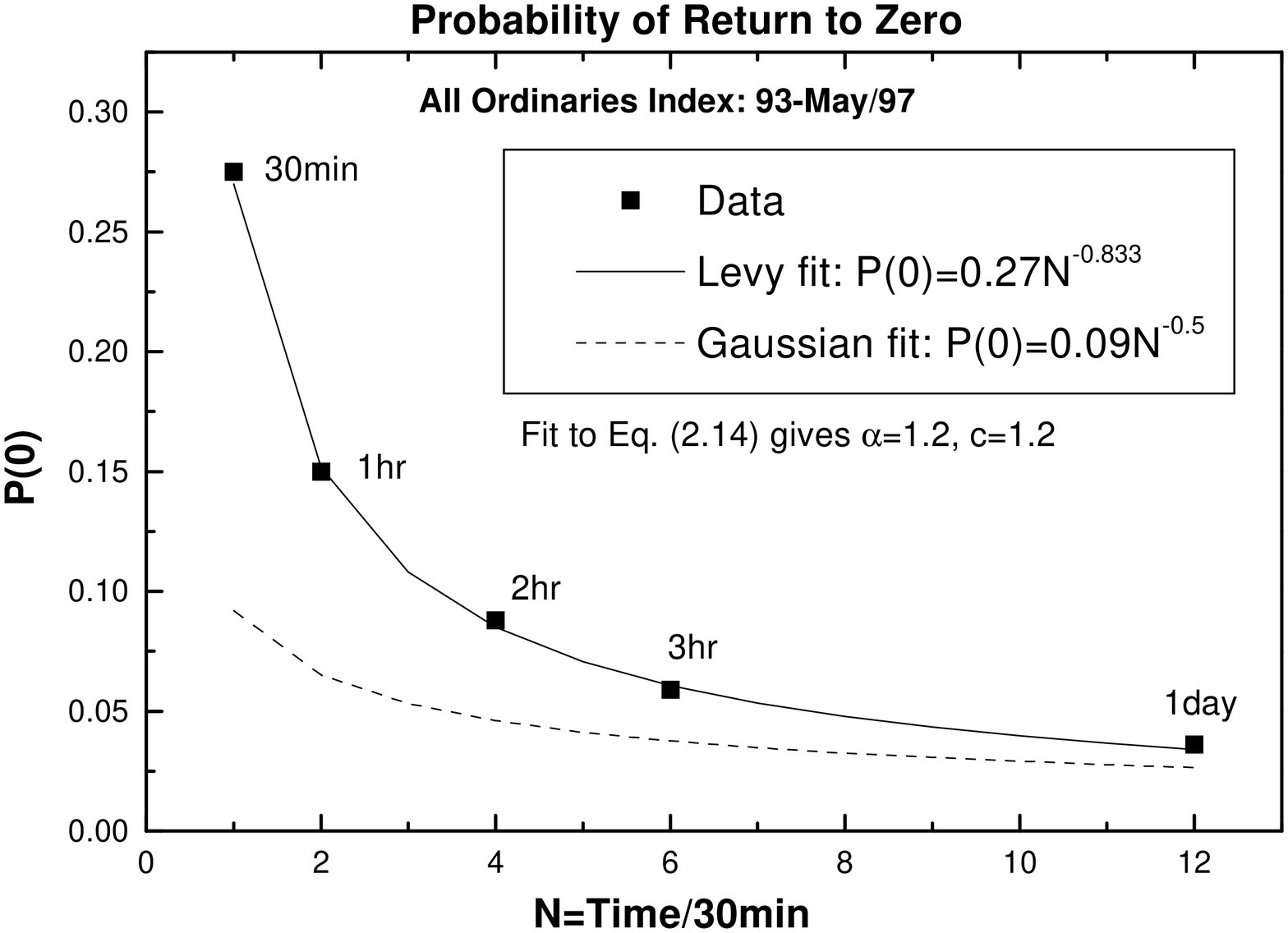}}
\vspace{-1.0cm}
\caption{Fit of Eq. (2.14) to high frequency data}
\end{figure}
\begin{figure}[tbp]
\epsfxsize=14.0cm
\epsfysize=10.0cm
\centering{\ \epsfbox{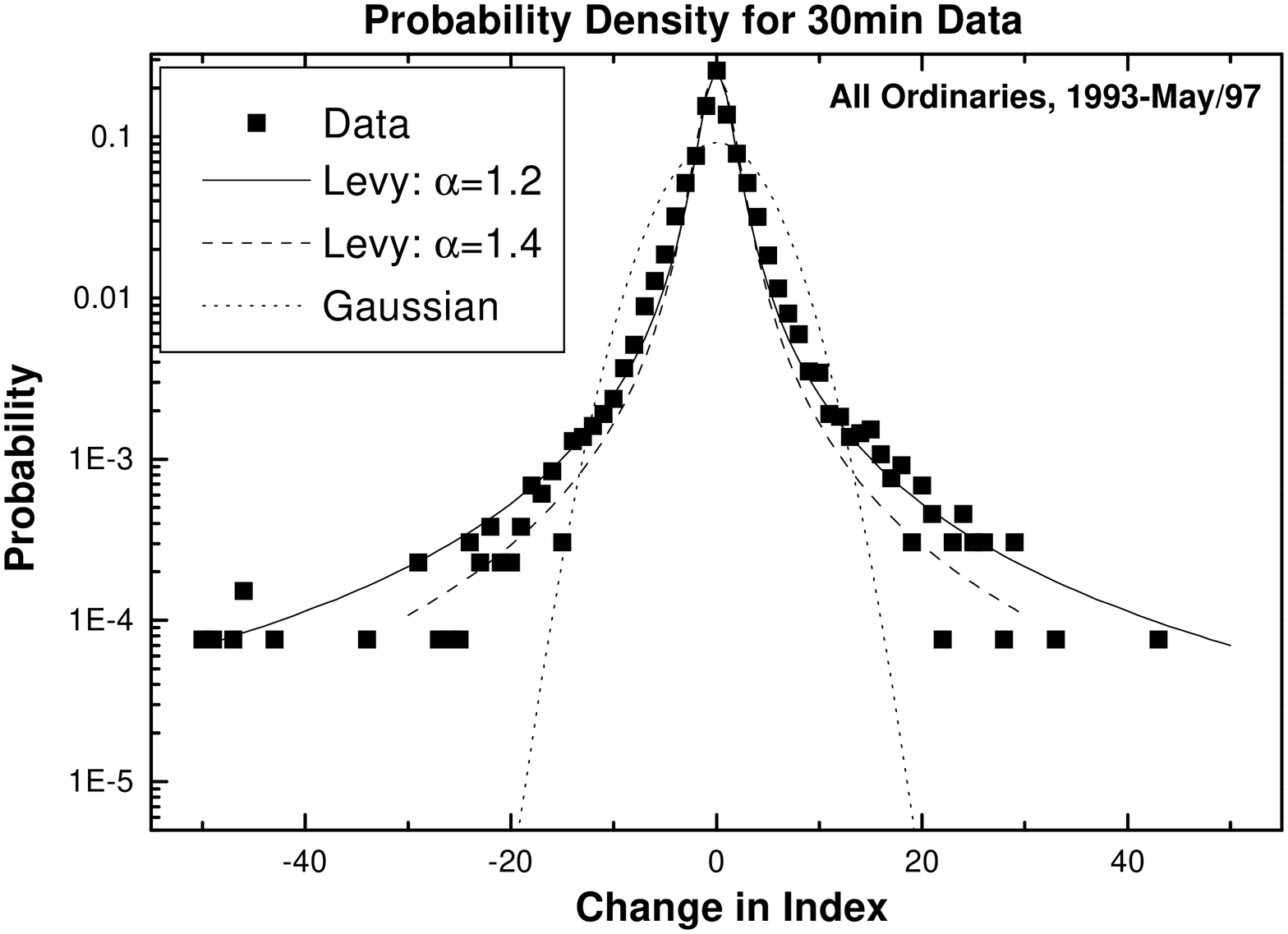}}
\vspace{-1.0cm}
\caption{Fit of the Levy PDF to 30 minute data using parameters derived 
from figure 1}
\end{figure}
\begin{figure}[tbp]
\epsfxsize=14cm 
\epsfysize=10.0cm
\centering{\ \epsfbox{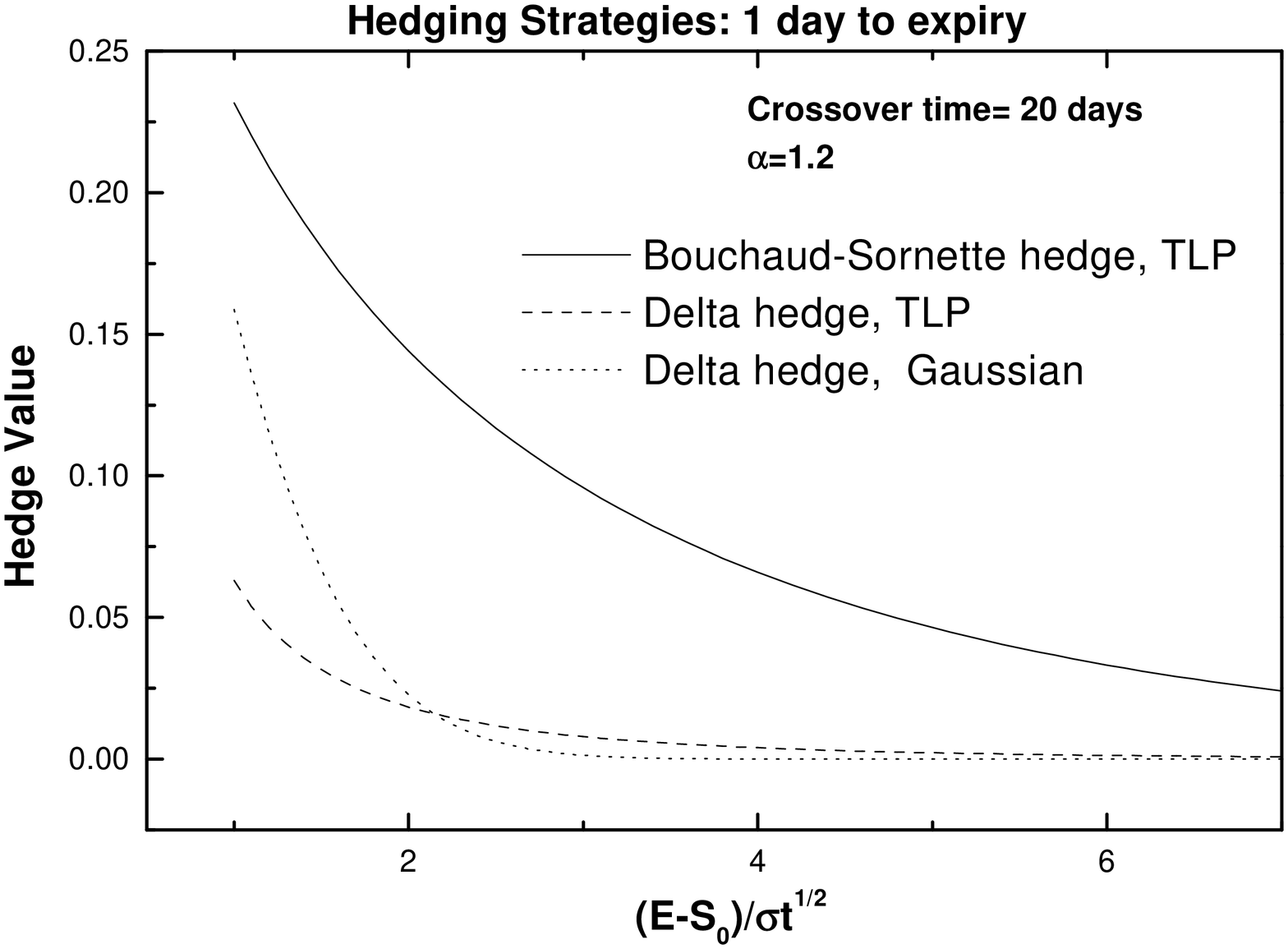}}
\vspace{-1.0cm}
\caption{Plot of the the TLP Bouchaud-Sornette optimal 
hedge (3.9), the TLP delta hedge (3.13) and the Gaussian delta hedge (3.6).}
\end{figure}
\begin{figure}[tbp]
\epsfxsize=14cm
\epsfysize=10.0cm
\centering{\ \epsfbox{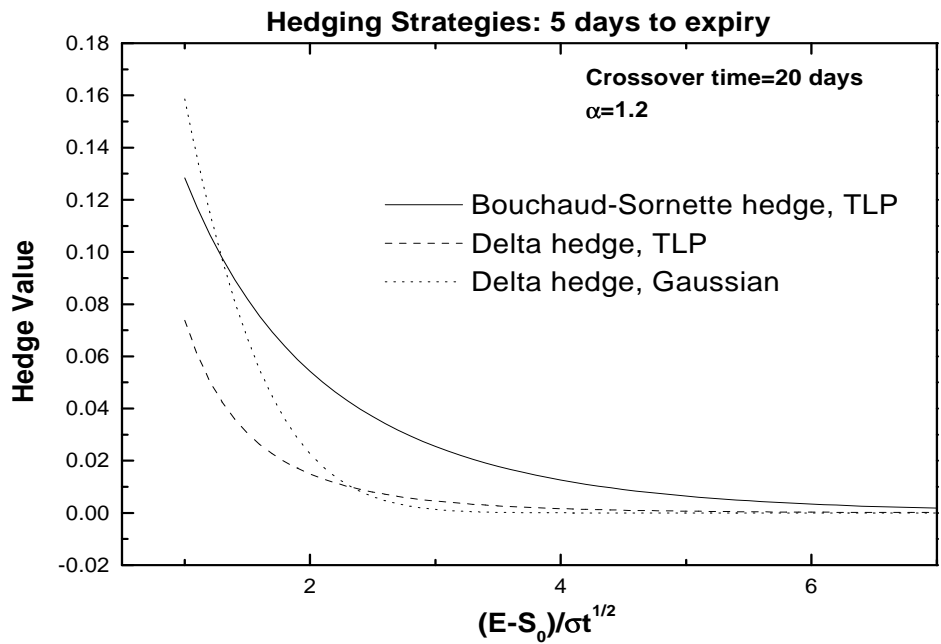}}
\vspace{-1.0cm}
\caption{Same as figure 3 except 5 days to expiry}
\end{figure}
\begin{figure}[tbp]
\epsfxsize=14cm
\epsfysize=10.0cm
\centering{\ \epsfbox{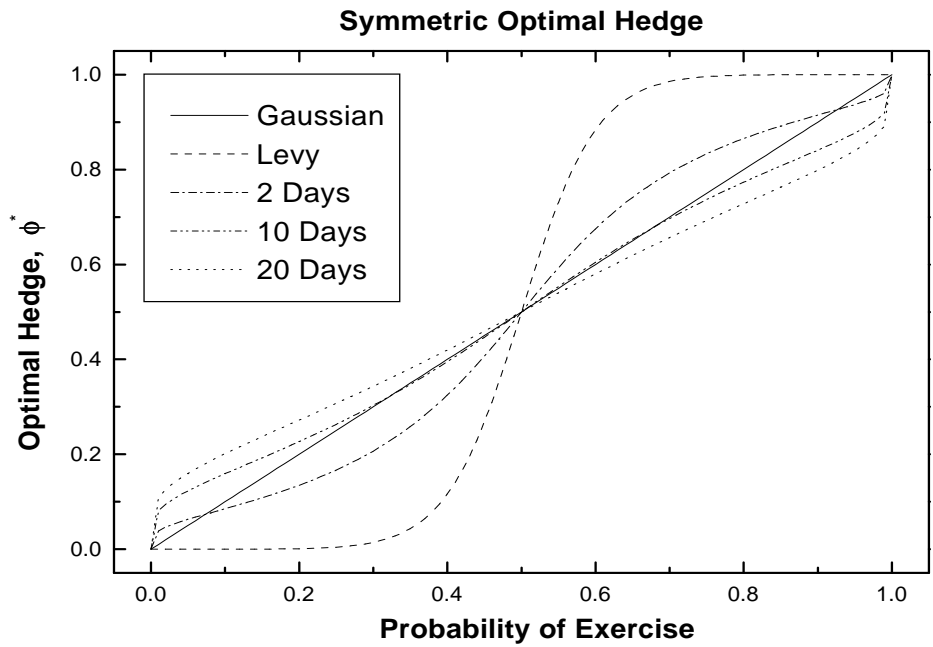}}
\vspace{-1.0cm}
\caption{Solution to Eq. (3.20) with $\a=1.2$ and $\lambda_-=\lambda_+=1/80$.}
\end{figure}
\begin{figure}[tbp]
\epsfxsize=14cm
\epsfysize=10.0cm
\centering{\ \epsfbox{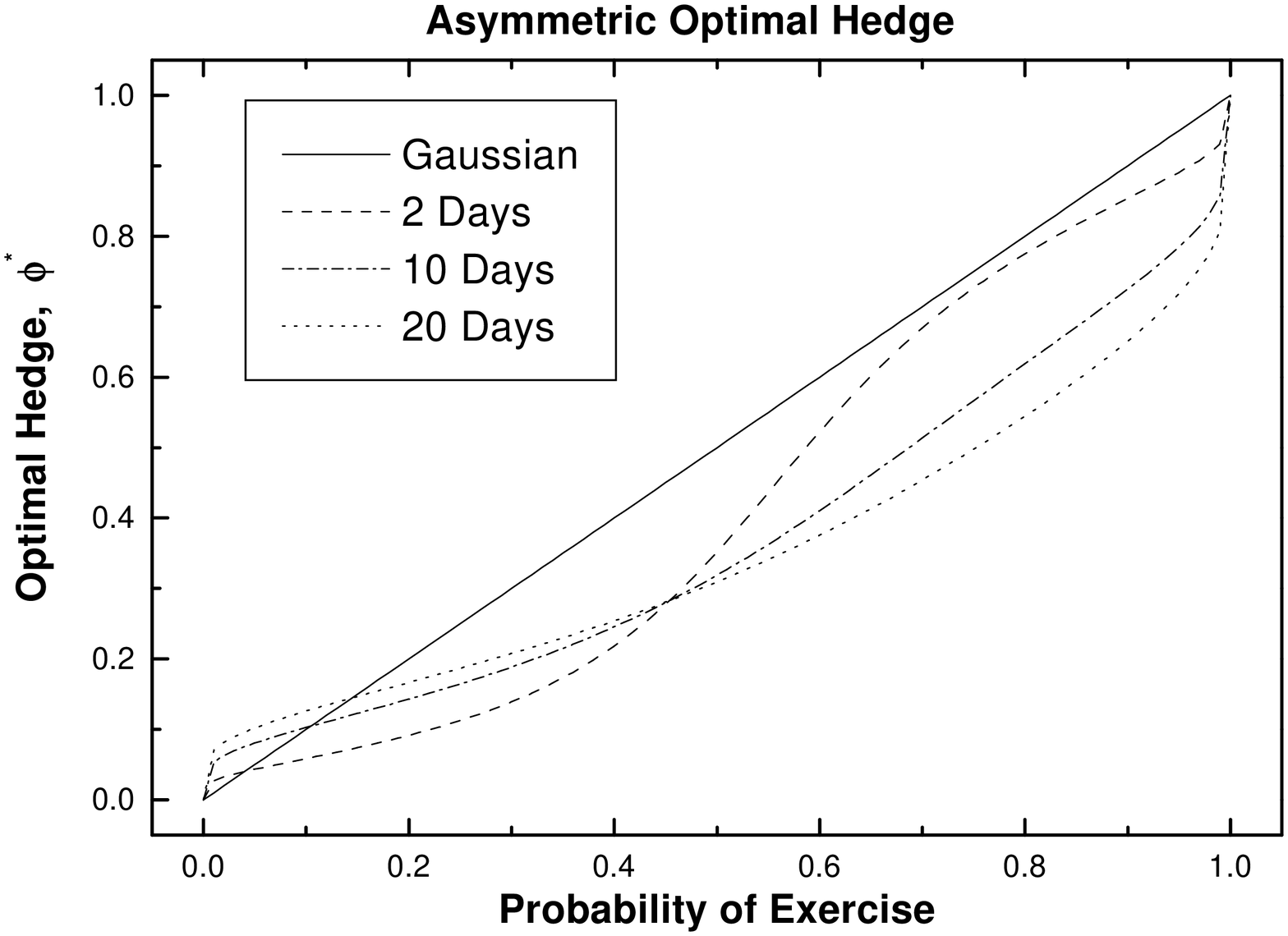}}
\vspace{-1.0cm}
\caption{Solution to Eq. (3.20) with $\a=1.2$, $\lambda_-=1/122$ and 
$\lambda_+=1/41$.}
\end{figure}


\begin{thebibliography}{999}

\bibitem{levy-empirical}
B.B. Mandelbrot, Journal of Business {\bf 36}, 394 (1963); 
B.B. Mandelbrot, Journal of Business {\bf 40}, 394 (1967); 
E. Fama, Management Science {\bf 11}, 404 (1965); J. H. McCulloch, Journal of 
Business {\bf 51}, No 4, (1978);
R.N. Mantegna, Physica A {\bf 179}, 232 (1991); S.T. Rachev and S. Mittnik, 
Econometric Reviews {\bf 12}, 261 (1993); 
E.E. Peters, {\it Fractal Market Analysis}, John Wiley and Sons, 1994.


\bibitem{levy-options}
D. Edelman, Abacus {\bf 31}, 113 (1995); J. P. Bouchaud, D. Sornette and 
M. Potters in {\it Proceedings of the Newton Institute session on Mathematical 
Finance} eds. M. Dempster and S. Pliska (Cambridge University Press, 1997). 

\bibitem{ak}
V. Akgiray and G.G. Booth, J. Business Econ. Statist. {\bf 6}, 51 (1988).

\bibitem{ms}
R.N. Mantegna and H.E. Stanley, Nature {\bf 376}, 46 (1995); 
R.N. Mantegna and H.E. Stanley, Physica A {\bf 239}, 225 (1997);
R.N. Mantegna in {\it The Physics of Complex Systems},  
Proceedings of the International School of Physics Enrico Fermi, Vol {\bf 134},
eds. F. Mallamace and H.E. Stanley (IOS Press, Amsterdam 1997).

\bibitem{hob} 
D.G. Hobson, {\it A Review of Stochastic Volatility Models} (1996),
available at: http://www.maths.bath.ac.uk/$\sim$oz/papers.html

\bibitem{vol}
R. Cont, preprint cond-mat/9705075; Y. Liu {\it et-al}, preprint 
cond-mat/9706021; P. Cizeau {\it et-al}, preprint cond-mat/9708143; 
A. Arneodo {\it et-al}, preprint cond-mat/9708012; B. Holdom, preprint 
cond-mat/9709141. \\ All available at: http://xxx.lanl.gov

\bibitem{tld-empirical}
R.N. Mantegna and H.E. Stanley, Nature {\bf 383}, 587 (1996);
A. Arneodo {\it et al}, preprint cond-mat/9607120 at http://xxx.lanl.gov;
R. Cont {\it et al}, preprint cond-mat/9705087 at http://xxx.lanl.gov;
J. P. Bouchaud, D. Sornette 
and M. Potters in {\it Proceedings of the Newton Institute session on 
Mathematical Finance} eds. M. Dempster and S. Pliska (Cambridge University 
Press, 1997); E. Aurell, J.P. Bouchaud, M. Potters and K. Zyczkowski, 
European Financial Management, (in press 1997).

\bibitem{man}
R.N. Mantegna and H.E. Stanley, Phys. Rev. Lett {\bf 73}, 2946 (1994);
R.N. Mantegna and H.E. Stanley in {\it Levy Flights and Related Topic in
Physics} eds. M.F. Shlesinger, G.M. Zaslavsky and U. Frisch (Springer, Berlin, 
1995). 

\bibitem{kop}
I. Koponen, Phys. Rev. E {\bf 52}, 1197 (1995).

\bibitem{bouchaud1}
J.P. Bouchaud and D. Sornette, J. Phys. I France {\bf 4}, 863 (1994); 
J.P. Bouchaud and D. Sornette, J. Phys. I France {\bf 5}, 219 (1995).

\bibitem{bouchaud2}
J.P. Bouchaud, G. Iori and D. Sornette, Risk {\bf 9(3)}, 61 (1996).

\bibitem{lev}
V. Zolotarev, {\it One-Dimensional Stable Distributions}, American Mathematical
Society, Providence RI (1986); J.P. Bouchaud and A. Georges, Physics Reports 
{\bf 195}, 125 (1990). 

\bibitem{shl}
M.F. Shlesinger, Phys. Rev. Lett {\bf 74}, 4959 (1995).

\bibitem{bs}
F. Black and M. Scholes, Journal of Political Economy {\bf 81}, 635 (1973).

\bibitem{hull}
see for eg. J. Hull, {\it Options, Futures, and other Derivatives} 3rd ed., 
Prentice Hall, NJ (1997).

\bibitem{sor}
D. Sornette, J. Phys. I France {\bf 7}, 1155-1171 (1997). 

\bibitem{sim}
J.M. Chambers, C.L. Mallows and B. Stuck, J. Amer. Statist. Assoc. {\bf 71},
340 (1976); R. Weron, Statist. Probab. Lett {\bf 28}, 165 (1996); 
R.N. Mantegna, Phys. Rev. E {\bf 49}, 4677 (1994).

\bibitem{mult}
S. Redner, Am. J. Phys. {\bf 58}, 267 (1990); J.M. Deutsch, Physica A {\bf
208}, 433 (1994).

\end{thebibliography}
\end{document}